\documentclass[preprint,showpacs,preprintnumbers,amsmath,amssymb]{revtex4}
\usepackage{graphicx}
\usepackage{dcolumn}
\usepackage{bm}

\newcommand{\be}{\begin{equation}}
\newcommand{\en}{\end{equation}}
\newcommand{\bea}{\begin{eqnarray}}
\newcommand{\ena}{\end{eqnarray}}

\begin{document}


\title{Intermediate inflation on the brane  }

\author{Sergio del Campo}
 \email{sdelcamp@ucv.cl}
\affiliation{ Instituto de F\'{\i}sica, Pontificia Universidad
Cat\'{o}lica de Valpara\'{\i}so, Casilla 4059, Valpara\'{\i}so,
Chile.}
\author{Ram\'on Herrera}
\email{ramon.herrera@ucv.cl} \affiliation{ Instituto de
F\'{\i}sica, Pontificia Universidad Cat\'{o}lica de
Valpara\'{\i}so, Casilla 4059, Valpara\'{\i}so, Chile.}
\date{\today}

\begin{abstract}
 Brane inflationary universe model in the context of intermediate inflation
  is studied. General
 conditions  for this model to be realizable are
 discussed. In the high-energy limit  we
 describe in great details the characteristic of this model.
  \end{abstract}

\maketitle

\section{Introduction}

It is well known that inflation is to date the most compelling
solution to many long-standing problems of the Big Bang model
(horizon, flatness, monopoles, etc.) \cite{guth,infla}. One of the
success of the inflationary universe model is that it provides a
causal interpretation of the origin of the observed anisotropy of
the cosmic microwave background (CMB) radiation, and also the
distribution of large scale structures \cite{astro,astro2}.

In concern to higher dimensional theories, implications of
string/M-theory to Friedmann-Robertson-Walker (FRW) cosmological
models have recently attracted  great deal of attention, in
particular, those related to brane-antibrane configurations such
as space-like branes\cite{sen1}. The realization  that we may live
on a so-called brane embedded in a higher-dimensional Universe has
significant implications to cosmology \cite{1}. In this scenario
the  standard model of particle is confined to the  brane, while
gravitation propagate into the bulk space-time. The effect of
extra dimensions induces additional terms in the Friedmann
equation \cite{3,8}. One of the term that appears in this
effective equation is  a   term proportional to  the square energy
density. This kind of models has been studied
extensively\cite{4,5}. For a review see Ref.\cite{M}.

On the other hand, intermediate inflation model was introduced as
an exact solution for a particular scalar field potential of the
type $V(\phi)\propto \phi^{-4(f^{-1}-1)}$\cite{Barrow1}, where $f$
is a free parameter. With this sort of potential, and with $1>f>
0$, it is possible in the slow-roll approximation to have a
spectrum of density perturbations which presents a scale-invariant
spectral index $n_s=1$, i.e. the so-called Harrizon-Zel'dovich
spectrum of density perturbations, provided $f$ takes the value of
two third\cite{Barrow2}. Even though  this kind of spectrum is
disfavored by the current WMAP data\cite{astro,astro2}, the
inclusion of tensor perturbations, which could be present at some
point by inflation and parametrized by the tensor-to-scalar ration
$r$, the conclusion that $n_s \geq 1$ is allowed providing  that
the value of $r$ is significantly nonzero\cite{ratio r}. In fact,
in Ref. \cite{Barrow3} was shown that the combination $n_s=1$ and
$r>0$ is given by a version of the intermediate inflation  in
which the scale factor varies as $a(t)\propto e^{t^{2/3}}$ within
the slow-roll approximation.

The main motivation to study intermediate inflationary model
becomes from string/M theory also. This theory suggests that in
order to have a ghost-free action high order curvature invariant
corrections to the Einstein-Hilbert action must be proportional to
the Gauss-Bonnet (GB) term\cite{17}. GB terms arise naturally as
the leading order of the ® expansion to the low-energy string
effective action, where ® is the inverse string tension\cite{18}.
This kind of theory has been applied to possible resolution of the
initial singularity problem\cite{19}, to the study of Black- Hole
solutions\cite{20}, accelerated cosmological solutions\cite{21},
among others. In particular , very recently, it has been found
that for a dark energy model the GB interaction in four dimensions
with a dynamical dilatonic scalar field coupling leads to a
solution of the form $a = a_0 exp At^f$\cite{22} , where the
universe starts evolving with a decelerated exponential expansion.
Here, the constant $A$ becomes given by $A = \frac{2}{\kappa n}$
and $f = \frac{1}{2}$ , with $\kappa^2 = 8 \pi G$ and $n$ is a
constant. In this way, the idea that inflation , or specifically,
intermediate inflation, comes from an effective theory at low
dimension of a more fundamental string theory is in itself very
appealing. Thus, in brane universe models the effective theories
that emerge from string/M theory lead to a Friedmann Equation
which is proportional to the square energy density, on the one
hand, and an evolving intermediate scale factor, in addition, it
makes interesting to study their mixture by itself, i.e., an
intermediate inflationary universe model in a brane world
effective theory.

The outline of the paper is as follows. The next section presents
a short review of the brane-intermediate Inflationary phase.
Section \ref{sectpert} deals with the calculations of cosmological
perturbations in general term. Finally, in Sect.\ref{conclu} we
summarize our finding.

\section{ The brane-intermediate Inflationary phase. }

 We consider the five-dimensional brane scenario in which the
Friedmann equation is modified from its usual form in the
following way\cite{2,3}
\begin{equation}
H^2=\kappa\,\rho_\phi\left[1+\frac{\rho_\phi}{2\lambda}\right]+\frac{\Lambda_4}{3}+\frac{\xi}{a^4},
\label{eq1}\end{equation} where  $H=\dot{a}/a$ denotes the Hubble
parameter, $\rho_\phi$ represents the matter field confined to the
brane, $\kappa=8\pi G/3=8\pi/3m_p^2$, $\Lambda_4$ is the
four-dimensional cosmological constant and  the last term
represents the influence of the bulk gravitons on the brane, where
$\xi$ is an integration constant. The brane tension $\lambda$
relates the four and five-dimensional Planck masses via the
expression $m_p=\sqrt{3M_5^6/(4\pi\lambda)}$, and it is
constrained by nucleosynthesis to satisfies the inequality
$\lambda >$ (1MeV)$^4$ \cite{Cline}. In the following, we will
assume the high energy regime, i.e. $\rho_\phi\gg \lambda$. Also,
we will take that the four-dimensional cosmological constant to be
vanished, and once inflation begins, the last  term will rapidly
become unimportant, leaving us with the effective Equation

\begin{equation}
H^2\simeq\beta\,\rho_\phi^2, \label{HC}
\end{equation}
where $\beta$ corresponds to $\beta=\kappa/(2\lambda)$, with
dimension of $m_P^{-6}$.

We assume that the scalar field is confined to the brane, so that
its field equation has the standard form
\begin{equation}
\dot{\rho_\phi}+3H(\rho_\phi+P_\phi)=0\,\,\,\,\mbox{or
equivalently } \,\,\, \ddot{\phi}\,+3H \; \dot{\phi}+V'=0.
\label{key_01}
 \en
Here, $\rho_\phi=\frac{\dot{\phi}^2}{2}+V(\phi)$, and
$P_\phi=\frac{\dot{\phi}^2}{2}-V(\phi)$, where
   $V(\phi)=V$ is
the scalar  potential. The dots mean derivatives with respect to
the cosmological time and
 $V'=\partial V(\phi)/\partial\phi$. For convenience we use
 units in which $c=\hbar=1$.

Exact solution  can  be found for intermediate inflationary
universe models where  the scale factor, $a(t)$, expands as
follows
\begin{equation}
a(t)=\exp(\,A\,t^{f}).\label{at}
\end{equation}
Here $f$ is a constant parameter with range $0<f<1$, and $A$ is a
positive constant with dimension of $m_P^f$.

From Eqs.(\ref{HC}), (\ref{key_01}) and using Eq.(\ref{at}) we
obtain
\begin{equation}
\dot{\phi}^2=-\frac{\dot{H}}{3\,\beta^{1/2}\,H}=\frac{(1-f)}{3\,\beta^{1/2}\,t},\label{fi}
\end{equation}
and
\begin{equation}
V=\frac{1}{\beta^{1/2}}\,\left[H+\frac{\dot{H}}{6\,H}\right]=
\frac{6\,A\,f\,t^f-(1+f)}{6\,\beta^{1/2}\,t}.\label{pot}
\end{equation}
The exact solution for the scalar field $\phi$ with potential
$V(\phi)$ can be found from  Eq.(\ref{fi})
\begin{equation}
(\phi-\phi_0)^2=\frac{4\,(1-f)}{3\,\beta^{1/2}}\,t,\label{exf}
\end{equation}
where $\phi(t=0)=\phi_0$. Now, by using Eqs.(\ref{pot}) and
(\ref{exf}) the scalar potential as a function of  $\phi$, becomes
\begin{equation}
V(\phi)=
\frac{2\,(1-f)}{9\beta}\,\left[6\,f\,A\,\left(\frac{3\,\beta^{1/2}}{4\,(1-f)}\right)^f\,(\phi-\phi_0)^{2(f-1)}
-\frac{(1-f)}{(\phi-\phi_0)^2} \right].\label{potex}
\end{equation}

The Hubble parameter as a function of the inflaton field, $\phi$,
becomes
\begin{equation}
H(\phi)=f\;A\;\left(\frac{3\,\beta^{1/2}\,(\phi-\phi_0)^2}{4\,(1-f)}\right)^{f-1}.\label{HH}
\end{equation}

The form for the scale factor $a(t)$ expressed by Eq.(\ref{at})
also arises when we solve the field equations in the slow roll
approximation, with a simple power law scalar potential. Assuming
the set of slow-roll conditions, i.e. $\dot{\phi}^2 \ll V(\phi)$
and $\ddot{\phi}\ll 3H\dot{\phi}$, the potential given by
Eq.(\ref{pot}) reduces  to
\begin{eqnarray}
V(\phi)=B\;(\phi-\phi_0)^{-2(1-f)},\label{inf2}
\end{eqnarray}
where
$$
B=\left[\frac{3}{4\,(1-f)}\right]^{f-1}\;f\,A\,\beta^{(f-2)/2},
$$
with dimension of $m_P^{6-2f}$. Here, the first term of
Eq.(\ref{potex}) dominates $V$ at large value of $(\phi-\phi_0)$.
Note that this kind of potential does not present a minimum. Also,
the solutions for $\phi(t)$ and $H(\phi)$ corresponding to this
potential  are identical  to those obtained when the exact
potential, Eq.(\ref{potex}), is used.

We  should note that in the low energy scenario  the scalar
potential becomes $V(\phi)\propto\,\phi^{-4(1-f)/f}$
\cite{Barrow1}, instead of the that occur in the  high energy
scenario in which $V(\phi)\propto\,\phi^{-2(1-f)}$. Without loss
of generality $\phi_0$ can be taken to be zero.

Introducing the dimensionless slow-roll parameters \cite{4}, we
write
\begin{equation}
\varepsilon=-\frac{\dot{H}}{H^2}\simeq\frac{1}{3\,\beta}\,\frac{V'^2}{V^3}=
\frac{4\,(1-f)^2}{3\,\beta}\;\frac{1}{\phi^2\,V}=\frac{4\,(1-f)^2}{3\,\beta\;B}\;\phi^{-2f},\label{ep}
\end{equation}
and
\begin{equation}
\eta=-\frac{\ddot{\phi}}{H\,
\dot{\phi}}\simeq\,\frac{V''}{3\,\beta\,V^2}=\frac{2\,(1-f)\,(3-2f)}{3\,\beta\,B}\;\phi^{-2f}\label{eta}.
\end{equation}
Note that the ratio between  $\varepsilon$ and $\eta$ becomes
$\varepsilon/\eta=\frac{2(1-f)}{(3-2f)}$ and  thus $\eta$ is
always larger than $\varepsilon$, since $0<f<1$. Note, also, that
$\eta$ reaches unity before $\varepsilon$ does. In this way, we
may establish that the end of  inflation is governed by the
condition $\eta=1$ in place of $\varepsilon=1$. From this
condition we get for the scalar field, at the end of inflation the
value
\begin{equation}
\phi_{end}=\left[\frac{2\,(1-f)\,(3-2\,f)}{3\,\beta\,B}\right]^{1/2f}.
\label{al}
\end{equation}
During inflation we take $V(\phi)\gg \lambda=\frac{\kappa}{2
\beta}$, so we assume that at the end of inflation the term
quadratic in the density still dominates against the linear term
(see Eq. (\ref{eq1})), so that
$$
B\phi_{end}^{2(f-1)} \gg \frac{\kappa}{2 \beta},
$$
and from Eq. (\ref{al}), the restriction on the parameter from the
high energy regime becomes \begin{equation} \beta \gg
\frac{\kappa^2}{4} \frac{1}{(Af)^{2/f}}\left
[\frac{3-2f}{2}\right]^{2(1-f)/f}. \label{14}
\end{equation}
The number of e-folds at the end of inflation using
Eq.(\ref{exf}) is given by
\begin{equation}
N=\int_{t_*}^{t_{end}}\,H\,dt=A\,(t_{end}^f-t_*^f)=\,A\,\left[\frac{3\,\beta^{1/2}}{4(1-f)}\right]^f\,
(\phi_{end}^{2f}-\phi_*^{2f}).\label{N}
\end{equation}

  In the following, the subscripts  $*$ and $end$ are
used to denote  the epoch when the cosmological scales exit the
horizon and the end of  inflation, respectively.

\section{Perturbations\label{sectpert}}

In this section we will study the scalar and tensor perturbations
for our model. It was shown in Ref. \cite{PB} that the
conservation of the curvature perturbation, $\cal{R}$, holds for
adiabatic perturbations, irrespective of the form of the
gravitational equations. One has
${\cal{R}}=\psi+H\delta\phi/\dot{\phi}\simeq
(H/\dot{\phi})(H/2\pi)$, where $\delta\phi$ is the perturbation of
the scalar field $\phi$. For a scalar field the power spectrum of
the curvature perturbations  is given  in the slow-roll
approximation by the following expression
${\cal{P}_R}\simeq\left.\left(\frac{H^2}{2\pi\dot{\phi}}\right)^2\right|_{\,k=k_*}$
\cite{4}, that in our case it becomes

\begin{equation}
{\cal{P}_R}\,\simeq\,\frac{3}{4\pi^2}\;
\frac{\sqrt{\beta}\;f^4\,A^4}{(1-f)}\;t_*^{4f-3}=\frac{f^4\,A^4}{\pi^2}\,
\left[\frac{3\,\sqrt{\beta}}{4\,(1-f)}\right]^{4f-2}\;\phi_*^{2(4f-3)}
,\label{dp}
\end{equation}
where we have used Eqs.(\ref{fi}) and (\ref{exf}). Here, $k_*$ is
referred to $k=Ha$, the value when the universe scale crosses the
Hubble horizon  during inflation.

 From Eqs.(\ref{al}), (\ref{N})
and (\ref{dp}), we obtained a constraint for the $A$ parameter
given by
\begin{equation}
A=\frac{8^{(2f-1)/3}}{f}\;\left[\frac{{\cal{P}_R}\,\pi^2\,\;(1-f)\;[3+2f(N-1)]^{(3-4f)/f}}{3
\,\sqrt{\beta}}\right]^{f/3}.\label{A1}
\end{equation}


From this latter equation we can obtain the value of $A$ for a
given values of $\beta$ and $f$ parameters when $N$ and $P_{\cal
R}(k_*)$ is given. In particular, for $f=1/2$ and $\beta=10^{10}
m_p^{-6}$ we get $A=0.046 m_p^{1/2}$. For $p=2/3$ and $\beta=
10^{13} m_p^{-6}$ we have $A=0.0014 m_p^{2/3}$. Here, we have
taken $P_{\cal R}(k_*)\simeq 2.4\times 10^{-9}$  and $N=60$.


The scalar spectral index $n_s$ is given by $ n_s -1 =\frac{d
\ln\,{\cal{P}_R}}{d \ln k}$,  where the interval in wave number is
related to the number of e-folds by the relation $d \ln k(\phi)=-d
N(\phi)$. From Eq.(\ref{dp}), we get,  $n_s  \simeq\,
1\,-2(3\varepsilon-\eta)$,
 or equivalently
\begin{equation}
n_s  \simeq
1\,-\frac{2}{3\,\beta\,V}\;\left[\frac{3\,V'^2}{V^2}-\frac{V''}{V}\right]
=1-\frac{4\,(1-f)\,(3-4f)}{3\,\beta\,B}\;\phi^{-2f}.\label{nsa}
\end{equation}
Since $1>f>0$, we clearly see that the Harrison-Zel'dovich model,
i.e., $n_s=1$ occurs for $f=3/4$. For  $n_s>1$ we have  $f>3/4$,
and $n_s<1$ is for $f>3/4$.

Note that the spectrum, Eq.(\ref{nsa}), exhibits properties
analogous to that found in the standard intermediate
inflation\cite{Barrow3}. Here, just like the standard intermediate
inflation, the spectrum offers properties which are different of
those showed by chaotic, power-law, extended inflationary universe
models, where $n_s$ is less than one. As was mentioned above we
could get values for $n_s$ greater than one.

One of the interesting features of the five-year data set from
Wilkinson Microwave Anisotropy Probe (WMAP) is that it hints at a
significant running in the scalar spectral index $dn_s/d\ln
k=n_{run}$ \cite{astro,astro2}. From Eq.(\ref{nsa}) we get  that
the running of the scalar spectral index becomes

\begin{equation}
n_{run}=-\left(\frac{2\,V}{V'}\right)\,\;[3\,\varepsilon_{,\,\phi}-\eta_{,\,\phi}]
\;\varepsilon=\frac{16\,f\,(f-1)^2\,(4f-3)}{9\,\beta^2\,B^2}\;\phi^{-4f}.\label{dnsdk}
\end{equation}
In models with only scalar fluctuations the marginalized value for
the derivative of the spectral index is approximately $-0.03$ from
WMAP-five year data only \cite{astro}.

On the other hand, the generation of tensor perturbations during
inflation would produce  gravitational waves. This perturbations
in cosmology are more involved in our case,  since in brane-world
gravitons propagate in the bulk. The amplitude of tensor
perturbations was evaluated in Ref.\cite{t}, where
${\cal{P}}_g=24\kappa\,\left(\frac{H}{2\pi}\right)^2F^2(x)$. In
our case we get
\begin{equation}
{\cal{P}}_g =\frac{6}{\pi^2}\kappa f^2A^2t^{2(f-1)}F^2(x)=
\frac{6}{\pi^2}\kappa
f^2A^2\left[\frac{3\,\beta^{1/2}\,\phi^2}{4\,(1-f)}\right]^{2(f-1)}\,F^2(x),\label{ag}
\end{equation}
where $x=Hm_p\sqrt{3/(4\pi\lambda)}=3\,H\,m_p^2\beta^{1/2}/4\pi$
and
$$
F(x)=\left[\sqrt{1+x^2}-x^2\sinh^{-1}(1/x)\right]^{-1/2}.
$$
Here the function $F(x)$ appeared from the normalization of a
zero-mode. The spectral index $n_g$ is given by $
n_g=\frac{d{\cal{P}}_g}{d\,\ln
k}=-\frac{2x_{,\,\phi}}{N_{,\,\phi}\,x}\frac{F^2}{\sqrt{1+x^2}}$.

From expressions (\ref{dp}) and (\ref{ag}) we write  the
tensor-scalar ratio as
\begin{equation}
r(k)=\left(\frac{{\cal{P}}_g}{P_{\cal R}}\right) \simeq
\left.8\kappa\frac{(1-f)}{f^2A^2\sqrt{\beta}}t^{1-2f}\,F^2(x)=\frac{6\kappa}{f^2A^2}
\left[\frac{3\beta^{1/2}}{4(1-f)}\right]^{-2f}\phi^{2(1-2f)}F^2(x)\right|_{\,k=k_*}.
\label{Rk}\end{equation}

 From Eqs.(\ref{nsa}) and (\ref{Rk}) we
can write the relation between the tensor-scalar ratio $r$ and the
scalar spectral index $n_s$ as
\begin{equation}
r=\frac{6\,\kappa}{f^2\,A^2}\;\left[\frac{4(1-f)}{3\,\beta^{1/2}}\right]^{(2f^2-2f+1)/f}\;
\left[\frac{(3-4f)}{B\,(n_s-1)\,\beta^{1/2}}\right]^{(1-2f)/f}\;F^2(n_s).
\end{equation}

Combining  WMAP five-year data\cite{astro,astro2} with the Sloan
Digital Sky Survey  (SDSS) large scale structure surveys
\cite{Teg}, it is found an upper bound for $r$ given by
$r(k_*\simeq$ 0.002 Mpc$^{-1}$)$ <0.28\, (95\% CL)$, where
$k_*\simeq$0.002 Mpc$^{-1}$ corresponds to $l=\tau_0 k\simeq 30$,
with the distance to the decoupling surface $\tau_0$= 14,400 Mpc.
The SDSS  measures galaxy distributions at red-shifts $a\sim 0.1$
and probes $k$ in the range 0.016 $h$ Mpc$^{-1}$$<k<$0.011 $h$
Mpc$^{-1}$. The recent WMAP five-year results give the values for
the scalar curvature spectrum $P_{\cal R}(k_*)\simeq
2.4\times\,10^{-9}$ and the scalar-tensor ratio $r(k_*)<0.43
(95$\%$ CL)$. We will make use of these values  to set constrains
on the parameters of our  model.

In Fig.(\ref{rons}) we show the dependence of the tensor-scalar
ratio on the spectral index for different values of the parameter
$f$. From left to right $f$=1/2, 2/3 and 4/5. From
Ref.\cite{astro}, two-dimensional marginalized
 constraints (68$\%$ and 95$\%$ confidence levels) on inflationary parameters
$r$, the tensor-scalar ratio, and $n_s$, the spectral index of
fluctuations, defined at $k_0$ = 0.002 Mpc$^{-1}$. The five-year
WMAP data places stronger limits on $r$ (shown in blue) than
three-year data (grey)\cite{Spergel}. In order to write down
values that relate $n_s$ and $r$, we used Eqs.(\ref{dp}),
(\ref{nsa}) and (\ref{Rk}). Also we have used the WMAP value
$P_{\cal R}(k_*)\simeq 2.4\times 10^{-9}$, and the value
$\beta=10^{10}\,m_p^{-6}$ (or equivalently the brane tension
$\lambda=\kappa/(2\,\beta)\simeq 4.2\times 10^{-10}\,m_p^4$ ).
Note that for  any value of the parameter $f$, (restricted to the
range $1>f>0$), our model is well supported by the data. From
Eqs.(\ref{al}), (\ref{N}), (\ref{A1}) and (\ref{Rk}), we observed
numerically that for $f = \frac{1}{2}$, the curve $r = r(n_s)$
(see Fig. (\ref{rons})) for WMAP 5-years enters the 95$\%$
confidence region for $r\simeq 0.37$, which corresponds to the
number of e-folds, $N \simeq 164$. For $r \simeq 0.30$ corresponds
to $N \simeq 219$, in this way the model is viable for large
values of the number of e-folds. For $f=1/2$ and in the case when
we increase the value of the parameter $\beta$ (or equivalently,
we decrease the brane tension) from $10^{12}m_P^{-6}$ to
$10^{18}m_P^{-6}$ we observe that the new line becomes quite
similar.

In Fig.(\ref{graf3}) we represent the dependence of the running of
the scalar spectral index on the spectral index for different
values of the parameter $f$, i.e $f$=1/2, 2/3 and 4/5. In tis plot
and from Ref.\cite{astro}, two-dimensional marginalized limits for
the spectral index, $n_s$, defined at $k_*$ = 0.002Mpc$^{-1}$, and
the running of the index $dns/d ln k$ (marked $n_{run})$, in which
models with no tensor contribution, and with a tensor contribution
marginalized over, are shown. In particular, for $f = 4/5$ we
observed numerically that with a tensor contribution the curve
$n_{run} = n_{run}(ns)$ enters the 95 $\%$ confidence region for
$n_{run}\simeq 0.004$ which correspond to $N\simeq 7$ . If the
number of e-folding is 60, we observed that $n_{run}\simeq 6\times
10^{-6}$. From the same figure and taking $\beta\sim
10^{10}m_p^{-6}$, we see that the model works quite well when $f$
lies in the range $4/5>f>2/3$.
\begin{figure}[th]
\includegraphics[width=5.0in,angle=0,clip=true]{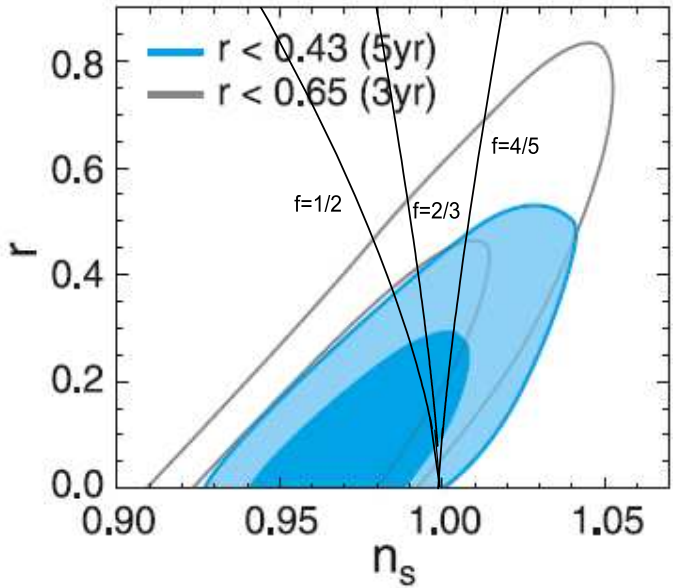}
\caption{The plot shows  $r$ versus $n_s$. Here, we have used the
WMAP five-year data where $P_{\cal R}(k_*)\simeq 2.4\times
10^{-9}$,  together with $\beta =10^{10} m_p^{-6}$.  The five-year
WMAP data places stronger limits on r (shown in blue) than
three-year data (grey)\cite{astro}.
 \label{rons}}
\end{figure}

\begin{figure}[th]
\includegraphics[width=5.0in,angle=0,clip=true]{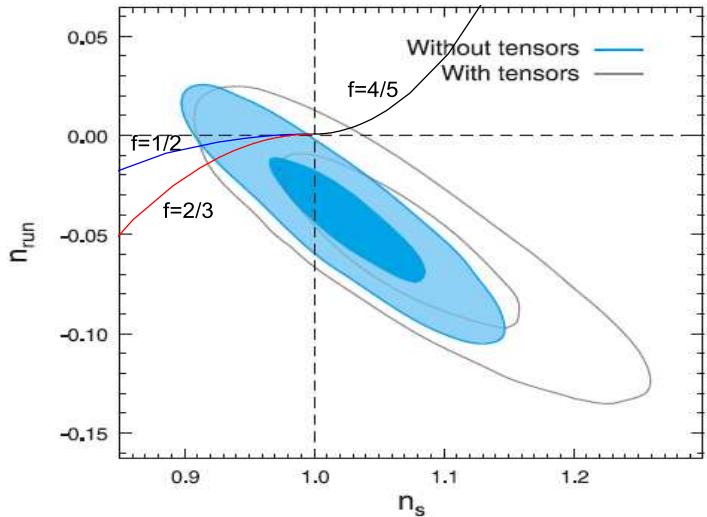}
\caption{The plot shows  $n_{run}=d\ln n_s/d\ln k$ versus $n_s$.
Here, we have used the WMAP five-year data where $P_{\cal
R}(k_*)\simeq 2.4\times 10^{-9}$,  together with the value $\beta
=10^{10} m_p^{-6}$. Models with no tensor contribution, and with a
tensor contribution marginalized over, are shown\cite{astro}.
 \label{graf3}}
\end{figure}


\section{Conclusions \label{conclu}}

In this paper we have studied the brane-intermediate inflationary
model in the high-energy scenario. We have found an exact solution
of the Friedmann equations for a flat universe containing a scalar
field $\phi(t)$ with potential $V(\phi)$. In the slow-roll
approximation we have found a general relation between the scalar
potential and its derivative. We have also obtained explicit
expressions for the corresponding, power spectrum of the curvature
perturbations $P_{\cal R}$, tensor-scalar ratio $r$, scalar
spectrum index $n_s$ and its running $n_{run}$.

By using  the scalar potential (see Eq.(\ref{inf2})) and from the
WMAP five year data,  we have found  constraints on the parameter
$A$  for a given values of $\beta$  and $f$. In order to bring
some explicit results we have taken the constraint $r-n_s$ plane
to first-order in the slow roll approximation. We noted that the
parameter $f$, which lies in the range $1>f>0$, the model is well
supported by the data as could be seen from Fig.(\ref{rons}).
Here, we have used the WMAP five year data, where $P_{\cal
R}(k_*)\simeq 2.4\times 10^{-9}$, and we have taken the value
$\beta=10^{10}m_p^{-6}$. On the other hand, Fig.(\ref{graf3})
clearly shows that for the value of $f>4/5$ allows $n_s>1$.
However, from Fig.(\ref{graf3}) the best values of $f$ occurs when
it lies in the range $4/5>f>2/3$.

In this paper, we have not addressed the phenomena of reheating
and possible transition to the standard cosmology  (see e.g.,
Refs.\cite{SRc,u,yo}). A possible calculation for the reheating
temperature in the hight-energy scenario would give new constrains
on the parameters of the model.  We hope to return to this point
in the near future.

\begin{acknowledgments}
  S.d.C. was
supported by COMISION NACIONAL DE CIENCIAS Y TECNOLOGIA through
FONDECYT grant N$^0$ 1070306. Also, from UCV-DGIP N$^0$ 123.787
(2008). R.H. was supported by the ``Programa Bicentenario de
Ciencia y Tecnolog\'{\i}a" through the Grant ``Inserci\'on de
Investigadores Postdoctorales en la Academia" \mbox {N$^0$
PSD/06}.
\end{acknowledgments}


\end{document}